\documentclass[twocolumn,showpacs,preprintnumbers,amsmath,amssymb]{revtex4}
\usepackage[dvipdfmx]{graphicx}
\usepackage{epsfig}

\newcommand{\GW}{$GW$}

\newcommand{\bfk}{{\bf k}}
\newcommand{\bfr}{{\bf r}}

\newcommand{\req}[1]{\mbox{Eq.~\!(\ref{#1})}}

\def\G0{G^0}

\def\QSGW{\rm QS{GW}}

\def\ei{\varepsilon_i}

\def\ej{\varepsilon_j}

\def\vxc{V^{\rm xc}}

\def\connect#1{\leavevmode{\setbox1=\hbox{#1}\copy1%
\raise .2\ht1 \vbox{\moveleft \wd1\vbox{\hrule width \wd1 height .5pt depth 0pt}}%
}}

\def\ftn[#1]{\rlap{\footnotemark[#1]}}

\def\H0{H^0}

\begin{document}

\title{Accurate energy bands calculated by the hybrid quasiparticle 
self-consistent $GW$ method implemented in the ecalj package}

\author{Daiki Deguchi}
\affiliation{Division of Materials and Manufacturing Science, Graduate School of Engineering, 
Osaka University, Suita, Osaka 565-0871, Japan}

\author{Kazunori Sato}
\affiliation{Division of Materials and Manufacturing Science, Graduate School of Engineering, 
Osaka University, Suita, Osaka 565-0871, Japan}
\email{E-mail: ksato@mat.eng.osaka-u.ac.jp}

\author{Hiori Kino}
\affiliation{National Institute for Materials Science, Tsukuba, Ibaraki 305-0047, Japan }

\author{Takao Kotani}
\affiliation{Department of Applied Mathematics and Physics, Tottori University, Tottori 680-8552, Japan}

\begin{abstract}
We have recently implemented a new version of the quasiparticle self-consistent
$GW$ (\QSGW) method in the ecalj package 
released at http://github.com/tkotani/ecalj. 
Since the new version of the ecalj package is numerically stable 
and more accurate than the previous versions, 
we can perform calculations easily without being
bothered with tuning input parameters.
Here we examine its ability to describe energy band properties, {\it e.g.}, 
band-gap energy, eigenvalues at special points, and effective mass, for a variety of 
semiconductors and insulators. 
We treat C, Si, Ge, Sn, SiC (in 2H, 3C, and 4H structures), 
(Al, Ga, In)$\times$(N, P, As, Sb), (Zn, Cd, Mg)$\times$(O, S, Se, Te),
SiO$_2$, HfO$_2$, ZrO$_2$, SrTiO$_3$, PbS, PbTe, MnO, NiO, and HgO.
We propose that a hybrid \QSGW\ method, where we mix 80\% of \QSGW\ and 
20\% of LDA, gives universally good
agreement with experiments for these materials.
\end{abstract}

\maketitle

\section{Introduction}
The quasiparticle self-consistent $GW$ (\QSGW) method 
is the best available method for finding a static one-body Hamiltonian 
$H^0$ that describes a system on the basis of the optimum independent-particle 
[or the quasiparticle (QP)] picture
\cite{Faleev04,van_schilfgaarde_quasiparticle_2006,kotani_quasiparticle_2007,shishkin_accurate_2007,kotani_quasiparticle_2014,klimes_predictive_2014,di_valentin_quasiparticle_2014}.
\QSGW\ considers for an optimum division of the full many-body Hamiltonian 
$H$ into $H=H^0+(H-H^0)$, by choosing $H^0$ so as to minimize 
the perturbative corrections caused by $(H-H^0)$ to the QPs described by $H^0$. 
That is, we perform a self-consistent calculation until the correction is minimized.
Note that $(H-H^0)$ should contain not only a bare 
Coulomb interaction but also a quadratic term,
which is missing in usual model Hamiltonians.
In the \QSGW\ methodm, we evaluate $(H-H^0)$ in the $GW$ approximation; therefore, 
in the determination of $H^0$, charge fluctuations 
(not only local fluctuations but also plasmons) are 
taken into account self-consistently within the random phase approximation (RPA).
We determine $H^0$, the self-energy $\Sigma(\bfr,\bfr',\omega)$,
and the effective screened Coulomb interaction $W(\bfr,\bfr',\omega)$
simultaneously when we attain self-consistency in \QSGW.

Kotani and his collaborators developed an all-electron \GW\ method 
based on the full-potential linear muffin-tin orbital (FPLMTO) 
method \cite{methfessel_electronic_2000} to perform \QSGW\ calculations 
\cite{Faleev04,kotani_quasiparticle_2007}. We call this method FPLMTO-\QSGW.
FPLMTO-\QSGW\ was applied to a wide range of materials and proved its potential 
to go beyond the abilities of current first-principles methods 
based on the density functional theory. 
\cite{kotani_quasiparticle_2007,di_valentin_quasiparticle_2014}.
However, mainly because of the difficulty in the usage of FPLMTO, it is 
too complicated to apply FPLMTO-\QSGW\ to a wide variety of materials.
The main reason for this is that FPLMTO uses only atom-centered localized
orbitals, Muffin-tin orbitals (MTOs), as a basis set to expand
eigenfunctions. Choosing parameters specifying MTOs is not straightforward and 
requires fine tunings and repeated tests to perform reliable calculations. 
In addition, the offset-$\Gamma$ method for the Brillouin zone (BZ) integration
is slightly problematic in treating anisotropic systems.

To avoid these problems, Kotani and coworkers have developed a new method of 
implementing the \QSGW\ method based on 
the linearized augmented plane-wave and muffin-tin orbital method (PMT method), which employs 
augmented plane wave (APW) and MTO basis sets \cite{kotani_quasiparticle_2014,kotani_formulation_2015}.
The PMT method is a unique mixed basis method that uses two types of augmented waves 
\cite{kotani_fusion_2010,kotani_linearized_2013,kotani_formulation_2015}.
We call this new implementation of \QSGW\ as PMT-\QSGW\ and 
the package is open for public use 
as the ecalj package, which is available at 
https://github.com/tkotani/ecalj/ \cite{ecalj}. 
In this paper, we present practical applications of PMT-\QSGW\ 
with the ecalj package to a variety of materials. 
The present calculation results have not been presented yet, 
although PMT-\QSGW\ with the ecalj package was already used in the preceding studies 
\cite{ryee2015,jang_quasiparticle_2015,han_quasiparticle_2014,nagara_band_2014,
geshi_pressure_2013,PSSB:PSSB201552490} and 
the minimum examination of PMT-\QSGW\ was reported in Ref. 5. 

\begin{table}[h]
\caption{\label{tab:mat}
Crystal structures used for the calculations in this paper and 
the number of $\bfk$ points for the self-energy calculations 
in the 1st Brillouin zone (see text). Labels mean as follows: 
COD: ID number of the crystal open database \cite{crytallographyopendatabase},
DIA: diamond structure, 
HEX: hexagonal structure, ZB: zincblende structure,
WZ: wurtzite structure, RS: rocksalt structure,
CUB: cubic structure, MONO: monoclinic structure,
TETRA: tetragonal structure, PERO: perovskite structure.}
\begin{tabular}[t]{l|lll}
\hline
\hline
        &  Lattice         & Crystal   & Number  \\
        &  constants (\AA) & structure & of ${\bf k}$ for $\Sigma$\\
\hline
C       &   $a=3.567$ & DIA & 6$\times$6$\times$6\\            
Si      &   $a=5.431$ & DIA & 6$\times$6$\times$6\\
Ge      &   $a=5.646$ & DIA & 6$\times$6$\times$6\\
Sn      &   $a=6.489$ & DIA & 6$\times$6$\times$6\\
\hline
SiC(2H) &  $a=3.076,c=5.048$ & HEX, & 6$\times$6$\times$3\\
                    &  & COD9008875 \\
SiC(3C) &  $a=4.348$ & ZB, &  6$\times$6$\times$6 \\
                   &  & COD9008856  \\
SiC(4H) &  $a=2.079,c=10.07$         &  HEX, Ref. 20 & 4$\times$4$\times$2\\
\hline
AlN(WZ) &  $a=3.112,c=4.982$         &  WZ         &  6$\times$6$\times$3  \\
AlN(ZB) &  $a=4.38$                  &  ZB         &  6$\times$6$\times$6  \\
AlP     &  $a=5.467$                 &  ZB         &  6$\times$6$\times$6  \\
AlAs    &  $a=5.661$                 &  ZB         &  6$\times$6$\times$6  \\
AlSb    &  $a=6.136$                 &  ZB         &  6$\times$6$\times$6  \\
\hline
GaN(WZ) &  $a=3.189,c=5.189$         &  WZ         &  6$\times$6$\times$3  \\
GaN(ZB) &  $a=4.50$                  &  ZB         &  6$\times$6$\times$6  \\
GaP     &  $a=5.451$                 &  ZB         &  6$\times$6$\times$6  \\
GaAs    &  $a=5.653$                 &  ZB         &  6$\times$6$\times$6  \\
GaSb    &  $a=6.096$                 &  ZB         &  6$\times$6$\times$6  \\
\hline
InN(WZ) &  $a=3.545,c=5.703$         &  WZ         &  6$\times$6$\times$3  \\
InN(ZB) &  $a=4.98$                  &  ZB         &  6$\times$6$\times$6  \\
InP     &  $a=5.870$                 &  ZB         &  6$\times$6$\times$6  \\
InAs    &  $a=6.058$                 &  ZB         &  6$\times$6$\times$6  \\
InSb    &  $a=6.479$                 &  ZB         &  6$\times$6$\times$6  \\
\hline
ZnO     &  $a=3.254$                 &  WZ         &  6$\times$6$\times$3  \\
ZnS(ZB) &  $a=5.413$                 &  ZB         &  6$\times$6$\times$6  \\
ZnS(WZ) &  $a=3.82,c=6.26$           &  WZ         &  6$\times$6$\times$3  \\
ZnSe    &  $a=5.667$                 &  ZB         &  6$\times$6$\times$6  \\
ZnTe    &  $a=6.101$                 &  ZB         &  6$\times$6$\times$6  \\
\hline
CdO     &  $a=4.72$                  &  RS         &  6$\times$6$\times$6  \\
CdS(ZB) &  $a=5.826$                 &  ZB         &  6$\times$6$\times$6  \\
CdS(WZ) &  $a=4.160,c=6.756$         &  WZ         &  6$\times$6$\times$3  \\
CdSe    &  $a=6.054$                 &  ZB         &  6$\times$6$\times$6  \\
CdTe    &  $a=6.482$                 &  ZB         &  6$\times$6$\times$6  \\
\hline
MgO     &  $a=4.212$                 &  RS         &  6$\times$6$\times$6  \\
MgS     &  $a=5.62$                  &  ZB         &  6$\times$6$\times$6  \\
MgSe    &  $a=5.91$                  &  ZB         &  6$\times$6$\times$6  \\
MgTe    &  $a=6.42$                  &  ZB         &  6$\times$6$\times$6  \\
PbS     &  $a=5.936$                 &  RS         &  6$\times$6$\times$6  \\
PbTe    &  $a=6.462$                 &  RS         &  6$\times$6$\times$6  \\
\hline
SiO$_2$c   &  $a=7.165$                 &  CUB        &  4$\times$4$\times$4  \\
HgO     &  $a=6.613,b=5.521$         &  MONO       &  2$\times$2$\times$4  \\
        &  $c=3.522$                 & COD9012530  &                       \\
ZrO$_2$    &  $a=3.559,c=5.111$         &  TETRA      &  4$\times$4$\times$2  \\
HfO$_2$    &  $a=3.545,c=5.102$         &  TETRA      &  4$\times$4$\times$2  \\
SrTiO$_3$  &  $a=3.90$                  &  PERO       &  4$\times$4$\times$4  \\
\hline		          
MnO     &  $a=4.445$ &  RS,AF-II, &  4$\times$4$\times$4 \\
        &            & Ref. 21 \\
NiO     &  $a=4.170$ & RS,AF-II,  &  4$\times$4$\times$4 \\
        &            & Ref. 21 \\
\hline
\hline
\end{tabular}

\end{table}

The purpose of the present study is twofold. On one hand, we show 
how accurately PMT-\QSGW\ describes the band structure of semiconductors and insulators. 
The first-principles calculations based on the local density approximation (LDA) are 
now frequently used to explain material properties, but owing to its deficiency in 
the prediction of band properties, such as energy gap and effective mass, 
the application range of the LDA is somewhat limited. Obviously, 
we cannot directly use the LDA to propose new photovoltaic or photocatalytic materials. 
Considering that the computational method becomes continuously important 
for exploring and fabricating new functional materials, 
it is important to show the reliability of the most advanced 
electronic structure theory, the so-called `beyond LDA' theory such as \QSGW, 
and encourage the first-principles 
calculations as a standard tool for designing new functional materials.  

The other purpose of the present study is to demonstrate the usability of the ecalj package.
Recently, the first-principles calculations have been used by both theoretical 
researchers and experimentalists for practical applications. 
In such a case, all users of the first-principles package are not always professionals. 
The ecalj package is designed so that 
all the calculations are performed essentially in its default settings; 
therefore, the users are not required to set the parameters that control 
the accuracy of the calculations. 
We only need to prepare information of crystal structures with a very limited number of inputs. 
By following the procedure described in Appendix, 
the users of the ecalj package can reproduce the results presented in this paper and 
actually observe the numerical stability and reliability of the ecalj package; thus 
the present paper serves a reference of standard calculation results. 

After the minimum explanation of PMT-\QSGW\ in the next section, 
we show the calculated band gaps. Then, we show the band properties and effective mass 
of materials with zincblende structures. Finally, we give a summary and
possible expectations for the PMT-\QSGW\ in the ecalj package.
We would conclude that the PMT-\QSGW\ in the ecalj package can be 
a useful tool for investigating problems not treated within the other 
standard electronic structure theories such as the LDA and hybrid methods. 
In Appendix, we show how to reproduce our results with the {\tt ecalj} package.

\begin{table*}[htb]
\begin{center}
\caption{\label{tab:bandgap}
Calculated minimum band-gap energies in eV by several calculation procedures. 
``+SO'' means that the spin-orbit interaction 
is included after the self-consistency is obtained. \QSGW80 means the calculation with 
80\% of \QSGW\ together with 20\% of LDA. 
\QSGW80(NoSC) means that we use \req{eq:vxchy} when we make the band plot after 
the convergence of (pure) \QSGW. 
\QSGW1shot means one-shot \QSGW\ (including off diagonal elements) from the LDA. 
In the LDA, we use the VWN exchange-correlation functional \cite{vwn}. Expt. means experimental values, 
and D/I distinguishes the direct or indirect band gap. Experimental values are taken from 
Ref. 21 otherwise indicated. \QSGW80+SO values together with 
LDA+SO and experimental ones are plotted in Fig.~\ref{fig:bandgap}.}
\begin{tabular}[t]{l|cccccccclc}
\hline
\hline
        &  LDA  &      LDA    &   QSGW    &  QSGW    &   QSGW  &  QSGW80   &  QSGW80  &QSGW80             &Expt.   &D/I \\  
        &       &        +SO  &           &    +SO    &   1shot &          &   +SO    &(NoSC)+SO                     \\
\hline
C       &  4.16   &  4.15       &  6.11       &  6.11       &  5.88       &  5.69       &  5.69       &  5.71       &  5.50       &  I  \\
Si      &  0.47   &  0.46       &  1.28       &  1.26       &  1.20       &  1.10       &  1.09       &  1.11       &  1.17       &  I  \\
Ge      &  0.00   &  0.00       &  1.03       &  0.93       &  0.81       &  0.80       &  0.70       &  0.74       &  0.79       &  I  \\
Sn      &  0.00   &  0.00       &  0.12       &  0.00       &  0.00       &  0.00       &  0.00       &  0.00       &  0          &  D  \\
\hline
SiC(2H) &  2.16   &  2.15       &  3.56       &  3.51       &  3.35       &  3.21       &  3.21       &  3.24       &  3.33       &  I  \\
SiC(3C) &  1.32   &  1.32       &  2.63       &  2.62       &  2.47       &  2.33       &  2.33       &  2.36       &  2.42       &  I  \\
SiC(4H) &  2.18   &  2.18       &  3.53       &  3.53       &  3.35       &  3.23       &  3.23       &  3.98       &  3.26\cite{PhysRev.133.A1163}       &  I  \\
\hline
AlN     &  4.34   &  4.34       &  6.91       &  6.91       &  6.41       &  6.30       &  6.29       &  6.40       &  6.19       &  D  \\
AlN(ZB) &  3.24   &  3.24       &  5.67       &  5.67       &  5.23       &  5.10       &  5.10       &  5.19       &  5.34\cite{thompson_deposition_2001}       &  I  \\
AlP     &  1.46   &  1.44       &  2.74       &  2.72       &  2.56       &  2.45       &  2.43       &  2.47       &  2.51       &  I  \\
AlAs    &  1.35   &  1.25       &  2.46       &  2.36       &  2.29       &  2.20       &  2.11       &  2.17       &  2.23       &  I  \\
AlSb    &  1.13   &  0.91       &  1.80       &  1.59       &  1.69       &  1.65       &  1.44       &  1.49       &  1.69       &  I  \\
\hline
GaN     &  1.91   &  1.91       &  3.84       &  3.83       &  3.45       &  3.38       &  3.38       &  3.45       &  3.50       &  D  \\
GaN(ZB) &  1.77   &  1.77       &  3.69       &  3.68       &  3.30       &  3.24       &  3.23       &  3.30       &  3.30\cite{ramirez-flores_temperature-dependent_1994}       &  D  \\
GaP     &  1.44   &  1.41       &  2.49       &  2.46       &  2.31       &  2.25       &  2.23       &  2.26       &  2.35       &  I  \\
GaAs    &  0.30   &  0.19       &  1.89       &  1.77       &  1.58       &  1.52       &  1.41       &  1.46       &  1.52       &  D  \\
GaSb    &  0.00   &  0.00       &  1.20       &  0.99       &  1.01       &  0.99       &  0.77       &  0.79       &  0.82       &  I  \\
\hline
InN     &  0.00   &  0.00       &  0.80       &  0.80       &  0.27       &  0.49       &  0.49       &  0.61       &  0.7\cite{PSSB:PSSB99991,PSSB:PSSB99994}        &  D  \\
InN(ZB) &  0.00   &  0.00       &  0.55       &  0.55       &  0.18       &  0.24       &  0.24       &  0.38       &  ---               &  D  \\
InP     &  0.46   &  0.43       &  1.65       &  1.62       &  1.40       &  1.37       &  1.34       &  1.38       &  1.42       &  D  \\
InAs    &  0.00   &  0.00       &  0.80       &  0.68       &  0.47       &  0.48       &  0.36       &  0.43       &  0.42       &  D  \\
InSb    &  0.00   &  0.00       &  0.77       &  0.54       &  0.51       &  0.49       &  0.25       &  0.29       &  0.24       &  D  \\
\hline
ZnO     &  0.74   &  0.72       &  3.88       &  3.87       &  2.91       &  3.10       &  3.10       &  3.26       &  3.44       &  D  \\
ZnS     &  1.86   &  1.83       &  4.12       &  4.10       &  3.62       &  3.57       &  3.55       &  3.65       &  3.71       &  D  \\
ZnS(WZ) &  1.94   &  1.92       &  4.21       &  4.18       &  3.70       &  3.66       &  3.64       &  3.74       &  3.91\cite{zakharov_quasiparticle_1994}       &  D  \\
ZnSe    &  1.06   &  0.93       &  3.23       &  3.10       &  2.73       &  2.71       &  2.58       &  2.68       &  2.82       &  D  \\
ZnTe    &  1.03   &  0.75       &  2.92       &  2.64       &  2.54       &  2.48       &  2.20       &  2.28       &  2.39       &  D  \\
\hline
CdO     &  0.00   &  0.00       &  1.32       &  1.32       &  0.53       &  0.85       &  0.84       &  0.98       &  1.09       &  I  \\
CdS     &  0.89   &  0.87       &  2.86       &  2.84       &  2.34       &  2.37       &  2.35       &  2.45       &  2.55\cite{zakharov_quasiparticle_1994}       &  D  \\
CdS(WZ) &  0.91   &  0.89       &  2.90       &  2.88       &  2.36       &  2.40       &  2.38       &  2.48       &  2.48       &  D  \\
CdSe    &  0.37   &  0.25       &  2.28       &  2.16       &  1.71       &  1.81       &  1.68       &  1.78       &  1.74       &  D  \\
CdTe    &  0.52   &  0.23       &  2.24       &  1.97       &  1.80       &  1.81       &  1.54       &  1.63       &  1.48       &  D  \\
\hline
MgO     &  4.77   &  4.76       &  8.97       &  8.96       &  8.22       &  7.98       &  7.97       &  8.14       &  7.67       &  D  \\
MgS     &  3.33   &  3.30       &  6.23       &  6.20       &  5.63       &  5.54       &  5.51       &  5.63       &  4.5        &  D  \\
MgSe    &  2.50   &  2.37       &  5.24       &  5.11       &  4.67       &  4.58       &  4.45       &  4.59       &  4.05       &  D  \\
MgTe    &  2.31   &  2.03       &  4.50       &  4.24       &  4.13       &  4.05       &  3.79       &  3.90       &  3.49       &  D  \\
PbS     &  0.26   &  0.07       &  0.73       &  0.49       &  0.63       &  0.62       &  0.36       &  0.39       &  0.19\cite{wei_electronic_1997}       &  D  \\
PbTe    &  0.67   &  0.02       &  1.06       &  0.47       &  0.99       &  0.98       &  0.36       &  0.38       &  0.29\cite{wei_electronic_1997}       &  D  \\
\hline
SiO$_2$(CUB)   &  5.43   &  5.43       &  10.09      &  10.09      &  9.29       &  9.05       &  9.05       &  9.19       &  8.9\cite{sio2cgap}       &  D  \\
HgO     &  1.11   &  1.09       &  2.89       &  2.89       &  2.53       &  2.49       &  2.46       &  2.54       &  2.8        &  I  \\
ZrO$_2$    &  3.84   &  3.84       &  6.83       &  6.83       &  6.12       &  6.07       &  6.07       &  6.11       &  5.68\cite{sayan_valence_2004}        &  I  \\
HfO$_2$    &  4.42   &  4.41       &  7.29       &  7.25       &  6.63       &  6.57       &  6.56       &  6.61       &  5.86\cite{sayan_band_2004}        &  I  \\
SrTiO$_3$  &  1.75   &  1.74       &  4.26       &  4.25       &  2.17       &  3.58       &  3.56       &  3.54       &  3.25\cite{van_benthem_bulk_2001}        &  I  \\
\hline
MnO     &  0.89   &  0.82       &  3.94       &  3.82       &  2.10       &  3.10       &  2.99       &  3.29       &  3.9\cite{tran_accurate_2009}        &  I  \\
NiO     &  0.59   &  0.59       &  5.59       &  5.59       &  2.16       &  5.29       &  4.54       &  5.00       &  4.3\cite{tran_accurate_2009}        &  I  \\
\hline
\hline
\end{tabular}
\end{center}
\end{table*}

\hspace{15cm}

\section{Method}
In this study, we apply the \QSGW\ method implemented in the {\tt ecalj} package to 
several materials. Readers are referred to Ref. 5 
for details on the theory and implementation of the \QSGW\ method. 
In this section, we give a minimum explanation of the \QSGW\ method.

In the LDA, we use $\vxc_{\rm LDA}(\bfr)$ calculated from the
electron density. This is calculated from a one-body Hamiltonian $H^0$. 
In contrast, we calculate $\Sigma(\bfr,\bfr',\omega)$ from eigenfunctions and
eigenvalues calculated from $H^0$. Then, we can
obtain the static but nonlocal exchange-correlation potential in \QSGW\
$\vxc_{\QSGW}(\bfr,\bfr')$, whose matrix elements are given as
\begin{eqnarray}
\vxc_{\rm QSGW} = \frac{1}{2}\sum_{ij} |\psi_i\rangle 
       \left\{ {{\rm Re}[\Sigma(\ei)]_{ij}+{\rm Re}[\Sigma(\ej)]_{ij}} \right\}
       \langle\psi_j|,\nonumber \\
\label{eq:vxc}
\end{eqnarray}
where $\ei$ and $|\psi_i\rangle$ are the eigenvalues and
eigenfunctions of $H_0$, respectively, and
$\Sigma_{ij}(\omega) = \langle \psi_i|\Sigma(\omega)| \psi_j \rangle =
\int d^3r \int d^3r' \psi_i^*(\bfr) \Sigma(\bfr,\bfr',\omega) \psi_j(\bfr')$.
${\rm Re}[\Sigma(\varepsilon)]$ is the real part of the self-energy, 
which assures the Hermiteness 
of the Hamiltonian \cite{Faleev04,kotani_quasiparticle_2007}.  With this $\vxc_{\rm QSGW}$, 
we can give a new static one-body Hamiltonian $H^0$ (by keeping 
$\vxc_{\rm QSGW}$ instead of using $\vxc_{\rm LDA}(\bfr)$, we run a
self-consistent calculation; then, the Hartree potential is also updated
since the electron density is updated).
Thus, we can perform the above procedure repeatedly until $H^0$ is converged. 
Simple semiconductors require about five iterations to achieve the convergence of 
eigenvalues within $\lesssim$ 0.01 eV. 
More iterations are required for materials such as antiferromagnetic NiO and MnO. 
We should emphasize the importance of off-diagonal elements of \req{eq:vxc}
in resolving band entanglement, {\it e.g.}, in Ge \cite{mark06adeq}.
In such a case, even for simple semiconductors such as Ge, 
we need fifteen iterations to obtain a well-converged band-gap energy. 

Generally, the \QSGW\ method systematically overestimates 
the band-gap energy \cite{van_schilfgaarde_quasiparticle_2006,kotani_quasiparticle_2007}. 
As suggested in Ref. 2, 
this can be due to the too small screening effect in the RPA, 
which neglects electron-hole correlations in the appropriate 
polarization function \cite{shishkin_accurate_2007},
and/or the screening effect of phonons suggested
by Botti and Marques \cite{botti_strong_2013}. 
Thus, it must be preferable to use the above-mentioned methods to remedy the
overestimation from the view point of physics; 
however, these methods can be computationally very demanding.
Instead, in this study, we use an empirical procedure, a hybrid \QSGW\ method 
introduced in Ref. 38. 
That is, we use
\begin{eqnarray}
\vxc = (1-\alpha) \vxc_{\rm QSGW} + \alpha \vxc_{\rm LDA},
\label{eq:vxchy}
\end{eqnarray}
where we assume $\alpha=0.2$, that is, 80\% \QSGW\ plus 20 \% LDA in
the calculations presented in this paper. We call this method \QSGW80. 
In this method, we use the $\vxc$ of \req{eq:vxchy} during the self-consistent
cycle of QSGW. In this study, we also tested `perturbative' \QSGW80, that is, 
we used the $\vxc$ of \req{eq:vxchy} after we obtained the self-consistent QSGW
results of $\alpha=0$. We call this `perturbative' \QSGW80 
`non-self-consistent \QSGW80', abbreviate as \QSGW80(NoSC) in the following.
This non-self-consistent procedure was previously used in Ref. 38. 
In Sect.~\ref{sec:result}, we note that \QSGW80 works reasonably well for a wide range of materials. 
This \QSGW80 can be a simple solution to treat interfaces or 
superlattices such as CdSe/CdS \cite{kocevski_first-principles_2015}, 
as long as both materials are described well with the same $\alpha$.

Spin-orbit coupling (SO) is essential for the 
correct prediction of effective mass and band-gap energy.
Considering the smallness of the effects of SO on the systems treated here, 
we can include SO as a perturbation 
after the scalar-relativistic self-consistent QSGW calculations.

\begin{figure*}[t]
\begin{center}
\includegraphics[width=16.5cm]{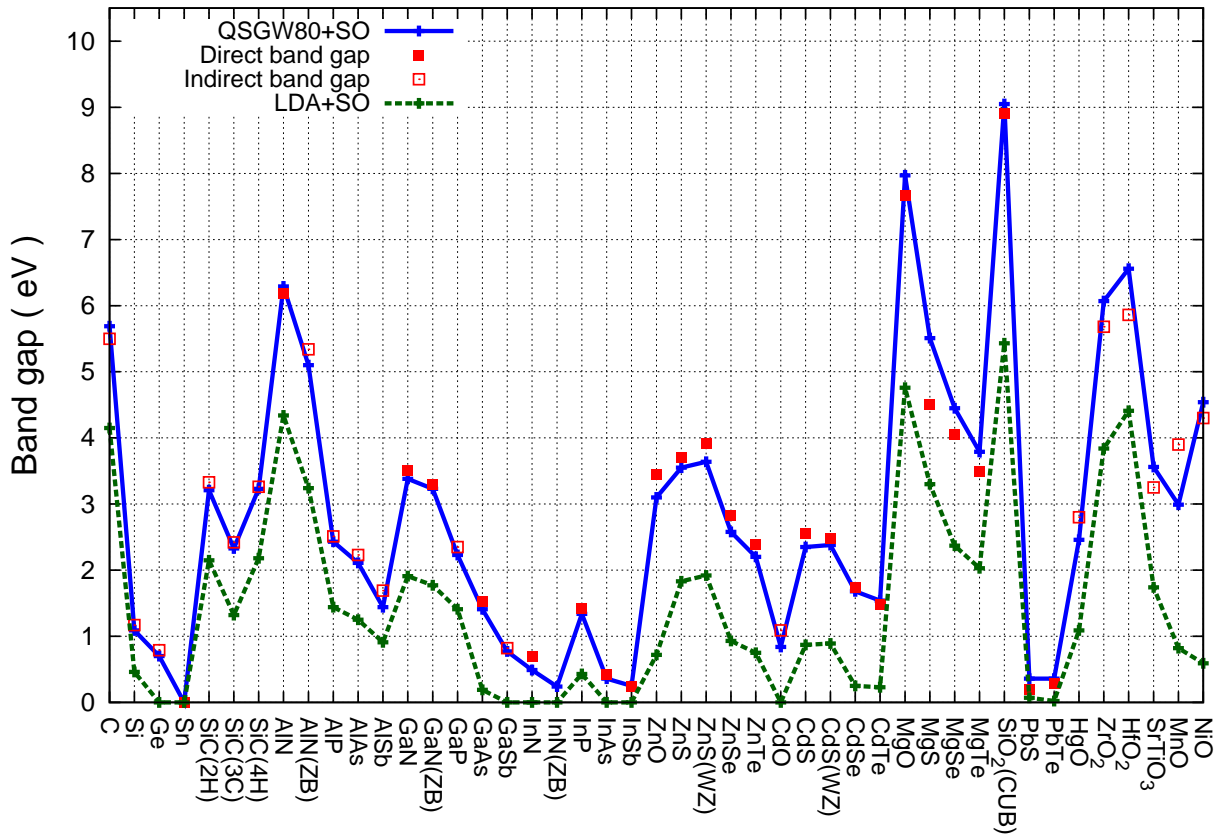}
\caption{\label{fig:bandgap}
Band-gap energies calculated by using \QSGW80+SO (blue solid line) and 
LDA+SO (green dotted line), together with experimental values 
(solid squares: direct band gap, open squares: indirect band gap). 
Respective values are shown in Table~\ref{tab:bandgap}.}
\end{center}
\end{figure*}

\section{Results}
\label{sec:result}
We calculate the properties of the materials shown 
in Table~\ref{tab:mat}. The crystal structures 
and lattice constants of the materials are taken from experimental values. 
In addition to the fundamental IV, III-V, and II-VI semiconductors,
we also include some important materials such as polytypes of SiC, Mg compounds, 
Pb compounds, and some oxides such as cubic SiO$_2$, HfO$_2$. 
We assume paramagnetic states except MnO and NiO, which exhibit a type II anti-ferromagnetic order
\cite{terakura84}. For some materials such as AlN, we treat two structures, both with 
zincblende (ZB) and wurtzite (WZ) structures.

As shown in Table~\ref{tab:mat}, the number of $\bfk$ points 
in the 1st Brillouin zone (BZ) for the
calculations of self-energy are $6 \times 6 \times 6$ for the zincblende structure in which 
two atoms are in the primitive cell. 
For the other structures, we reduce the number so as to keep the number of 
$\bfk$ points per atom almost the same, {\it e.g.}, we use $6 \times 6
\times 3$ for the wurtzite structure. 

In the ecalj package, we use an interpolation
technique for $\vxc_{\rm QSGW}$ in the entire BZ \cite{kotani_quasiparticle_2014}. 
This interpolation allows us to use a large number of $\bfk$ points 
in the step for determining $H^0$ for the given $\vxc_{\rm QSGW}$. 
$\vxc_{\rm QSGW}$ is calculated for the number of $\bfk$ points shown in Table~\ref{tab:mat}.

From the results of convergence check as shown in Ref. 5, 
we infer that 
numerical errors can be $\sim$ 0.1 eV owing to the settings of parameters 
in the calculations. The number of MTOs per atom was $\sim$30, and it
depends on atomic species. 
The cutoff energy of APWs is 3 Ry = 40.8 eV, which is good enough to
reproduce energy bands without empty spheres \cite{kotani_quasiparticle_2014}.

\begin{table}[t]
\caption{\label{tab:bandprop}
Eigenvalues (in eV) relative to the top of the valence band at the $\Gamma$ point 
for five selected zincblende materials. The values obtained by HSE06 \cite{kim_towards_2010} 
include the spin-orbit coupling. 
Experimental values are also taken from Ref. 40.
}
\begin{tabular}[t]{lccccc}
\hline
\hline
    Material &  $E_g$     &    \QSGW & \QSGW80  &  HSE06 & Expt. \\
             &            &     +SO &    +SO  &        &      \\
\hline
 InP  & $\Gamma_6^c$   &     1.62   &    1.34  &   1.48  &   1.42  \\
      & $X_6^c$   &     2.48   &    2.26  &   2.35  &   2.38  \\
      & $X_7^v$   &    -2.54   &   -2.49  &  -2.52  &  -2.20  \\
      & $L_6^c$   &     2.46   &    2.18  &   2.25  &   2.01  \\
      & $L_{4,5}$ &    -1.05   &   -1.03  &  -1.03  &  -1.00  \\
\hline
 InAs & $\Gamma_6^c$   &     0.68   &    0.36  &   0.42  &   0.42  \\
      & $X_6^c$   &     2.09   &    1.88  &   1.98  &   1.90  \\
      & $X_7^v$   &    -2.65   &   -2.60  &  -2.64  &  -2.70  \\
      & $L_6^c$   &     1.74   &    1.46  &   1.53  &   ----  \\
      & $L_{4,5}$ &    -1.07   &   -1.05  &  -1.06  &  -0.90  \\
\hline
 InSb & $\Gamma_6^c$   &     0.54   &    0.25  &   0.28  &   0.24  \\
      & $X_6^c$   &     1.55   &    1.41  &   1.53  &   1.80  \\
      & $X_7^v$   &    -2.62   &   -2.57  &  -2.66  &  -2.24  \\
      & $L_6^c$   &     1.00   &    0.79  &   0.85  &   0.93  \\
      & $L_{4,5}$ &    -1.11   &   -1.09  &  -1.12  &  -1.05  \\
\hline
 GaAs & $\Gamma_6^c$   &     1.77   &    1.41  &   1.33  &   1.52  \\
      & $X_6^c$   &     2.09   &    1.88  &   1.96  &   2.18  \\
      & $X_7^v$   &    -2.99   &   -2.93  &  -2.99  &  -2.80  \\
      & $L_6^c$   &     1.98   &    1.69  &   1.67  &   1.85  \\
      & $L_{4,5}$ &    -1.25   &   -1.23  &  -1.25  &  -1.30  \\
\hline
 GaSb & $\Gamma_6^c$   &     1.09   &    0.77  &   0.72  &   0.81  \\
      & $X_6^c$   &     1.19   &    1.05  &   1.26  &   1.14  \\
      & $X_7^v$   &    -2.90   &   -2.86  &  -2.95  &  -2.72  \\
      & $L_6^c$   &     0.99   &    0.78  &   0.87  &   0.88  \\
      & $L_{4,5}$ &    -1.67   &   -1.26  &  -1.29  &  -1.32  \\
\hline
\hline
\end{tabular}
\end{table}

\subsection{Minimum band gap}

The calculated minimum band-gap energies are shown in Table~\ref{tab:bandgap}.
The label ``+SO'' means that the SO coupling was added after the convergence of 
the \QSGW\ iteration. The label ``QSGW80'' means the hybrid calculation 
with $\alpha=0.2$ in \req{eq:vxchy}, that is, 80\% of \QSGW\ plus 20\% of LDA.
We observe that the normal \QSGW\ with SO, {\it i.e.}, \QSGW+SO, 
systematically overestimates the band-gap energy in comparison with experimental values. 
This overestimation was already observed in FPLMTO-QSGW 
\cite{van_schilfgaarde_quasiparticle_2006,kotani_quasiparticle_2007}. 
In contrast, we observe that \QSGW80+SO shows a much better agreement 
with experiments systematically.
In Fig.~\ref{fig:bandgap}, we plot the \QSGW80+SO values given in Table
\ref{tab:bandgap}, together with experimental values. 

For most of the semiconductors shown in Table~\ref{tab:bandgap}, 
the deviations of the theoretical predictions from the experimental values are 
as large as $\sim 0.1$ eV. This is in the numerical 
uncertainty range of our implementation. 
In some cases, error can be slightly larger. 
For example, the calculated energy gap of InN(WZ) is 0.49 eV, which is 0.21 eV 
smaller than the experimental value of 0.7 eV.
As for ZnO, the calculated value of 3.44 eV is 0.34 eV away from the experimental one of 3.10 eV. 
This is a case where the calculated band gap is largely affected by $\alpha$ 
since the LDA value is very small. For practical application to semiconductors, 
we need to take into consideration the accuracy shown in Table~\ref{tab:bandgap}.
On the other hand, we should note that the experimental error bar of band gaps of materials 
(especially oxides) may not be sufficiently small, for example, the energy gap of MnO is 
$3.9\pm0.4$ eV.

In \QSGW80(NoSC)+SO, we use \req{eq:vxchy} after we obtain the self-consistent $\vxc_{\rm QSGW}$
with $\alpha=0$ (usual \QSGW). The difference from \QSGW80+SO is not very 
large except in cases such as NiO, where \QSGW80(NoSC)+SO gives 5.00 eV and 
\QSGW80+SO gives 4.54 eV. 
This difference is due to the localized 3d electrons on cation sites. 
One may think that the large difference is due to the existence of 
the localized 3d orbital near the Fermi level, but it is not true because 
the energy gaps of 3.56 and 3.54 eV are obtained 
for SrTiO$_3$, electronic state of which is close to that of 
NiO apart from its magnetic nature. 
We can not discuss it further because we do not have enough data to 
investigate its origin. We observe smaller but slightly different values 
even in ZnO (3.10 and 3.26 eV).

We can observe the degree of the effect of SO as the difference between, 
{\it e.g.}, \QSGW\ and \QSGW+SO. As we see, 
materials including heavier atoms (especially anion) show larger SO effects.
The difference between \QSGW\ and \QSGW+SO is very similar to
that between \QSGW80 and \QSGW80+SO. 
This justifies our perturbative procedure for the SO coupling. 

The label ``QSGW1shot'' means the 1shot $GW$ calculation from the LDA result 
and the diagonalization including the off-diagonal elements of $\vxc_{\rm QSGW}$ in \req{eq:vxc}.
Thus, we can solve the entanglement problem. This is why our calculation 
gives the band gap for Ge, where the usual one-shot $GW$ (eigenvalue shift only) 
cannot give the band gap \cite{mark06adeq}.
In some cases, \QSGW1shot can be useful from the view point of computational speed.
\QSGW1shot can be used as a practical tool since it can give
a reasonable agreement with experiments for many semiconductors
(To compare the calculated values with experimental values, we need to add the SO 
effect, which can be taken as the difference between \QSGW80 and
\QSGW80+SO), although it is not applicable to materials such as NiO.

\begin{table}[t]
\caption{\label{tab:effmass}
Effective mass ($m_{\rm e}$: electron, $m_{\rm lh}$: light hole, 
$m_{\rm hh}$: heavy hole, and $m_{\rm so}$: split off band) for
zincblend materials along [100] direction calculated by \QSGW80+SO. 
See text and Fig.~\ref{fig:massgaas} about how to calculate these values.
We have poor fitting for $m_{\rm lh}$ and $m_{\rm so}$ 
for GaN(ZB) and InN(ZB); thus, no results are shown.
Values in parentheses are experimental values taken from Ref. 40.
}
\begin{tabular}[t]{l|cccccccc}
\hline
\hline
        &  $m_{\rm e}$   &  $m_{\rm lh}$  &  $m_{\rm hh}$  &  $m_{\rm so}$    \\
\hline
GaN(ZB) &  0.188   &  ----    &  0.807   &  ----   \\
GaP     &  0.128   &  .159    &  0.369   &  0.231  \\
GaAs    &  0.066   &  0.083   &  0.317   &  0.164  \\
        &  (0.067) &  (0.090) & (0.350)  & (0.172) \\
GaSb    &  0.043   &  0.048   &  0.232   &  0.143  \\
        & (0.039)  & (0.044)  &  (0.250) & (0.120) \\
InN(ZB) &  0.024   &  ---     &  1.022   &  ----    \\
InP     &  0.079   &  0.101   &  0.412   &  0.173  \\
        & (0.080)  & (0.121)  & (0.531)  & (0.210) \\
InAs    &  0.024   &  0.028   &  0.344   &  0.100  \\
        & (0.026)  &  (0.027) &  (0.333) & (0.140) \\
InSb    &  0.017   &  0.019   &  0.251   &  0.126   &  \\
        &  (0.014) & (0.015)  & (0.263)  & (0.110)  \\
ZnS     &  0.186     &  0.252     &  0.643     &  0.376    \\
ZnSe    &  0.128     &  0.175     &  0.542     &  0.313    \\
ZnTe    &  0.112     &  0.135     &  0.395     &  0.286    \\
CdS     &  0.152     &  0.200     &  0.698     &  0.331     \\
CdSe    &  0.105     &  0.143     &  0.578     &  0.286     \\
CdTe    &  0.093     &  0.114     &  0.420     &  0.285     \\
MgS     &  0.248     &  0.409     &  1.261     &  0.634     \\
MgSe    &  0.200     &  0.327     &  1.037     &  0.553     \\
MgTe    &  0.174     &  0.258     &  0.732     &  0.495     \\

\hline
\hline
\end{tabular}
\end{table}

\subsection{Band property of ZB-type semiconductors}
We sometimes need quasi-particle energies at specific ${\bf k}$ points in BZ, 
for example, when we observe optical responses, which contain both direct and indirect
transitions, or when we need to consider intervalley transitions.
Here, we show the eigenvalues instead of band plots (see
Appendix about how to show band plots).

In Table~\ref{tab:bandprop}, we show the eigenvalues, which can be interpreted 
as quasi-particle energies, at the $\Gamma, X$, and $L$ points for five selected ZB materials. These values can be compared with those calculated in
the hybrid functional HSE06 \cite{kim_towards_2010,heyd_hybrid_2003,heyd_erratum:_2006}.
We can consider that \QSGW80+SO shows good agreement with experiments. However, we
also see its limitations of accuracy.
For example, $L_6^c$ and $\Gamma_6^c$ for InP by \QSGW80+SO are 2.18 and 1.34 eV, 
respectively. They are slightly different from the experimental values of 2.01 ($L_6^c$) and 
1.42 ($\Gamma_6^c$) eV. The difference is slightly larger than the agreement 
of the band gap itself [1.34 eV (\QSGW80+SO) and 1.42 eV (experiment) at the $\Gamma$ point]. 
Thus, care should be taken about this level of error when we apply 
\QSGW80+SO to materials. However, we simultaneously need to note the accuracy 
of experimental data shown in Table~\ref{tab:bandprop}. 
These data are mostly estimated by optical experiments, 
where we need to remove excitonic effects theoretically from the raw experimental data 
based on some simple assumptions. Therefore, the difference between 
\QSGW80+SO and HSE06 could be within the experimental error range.

HSE06 values, taken from Ref. 40, 
also give good agreement with experimental values. 
HSE06 uses GGA, but corrects a short-range exchange 
part with the Hartree-Fock exchange term. 
Since the Hartree-Fock method can give very large band gaps, 
results can be strongly affected by the hybridization ratio. 
This is in contrast to \QSGW80 where we only hybridized 20\% of LDA so
as to correct the small error of \QSGW. One interesting observation is that
HSE06 and \QSGW80+SO give similar tendencies as for differences from
experimental values. For example, the valence band width $X_7^v$ values of InP  are 
-2.49 and -2.52 eV for \QSGW80+SO and HSE06, respectively.

\begin{figure*}[t]
\begin{center}
\includegraphics[width=16.5cm]{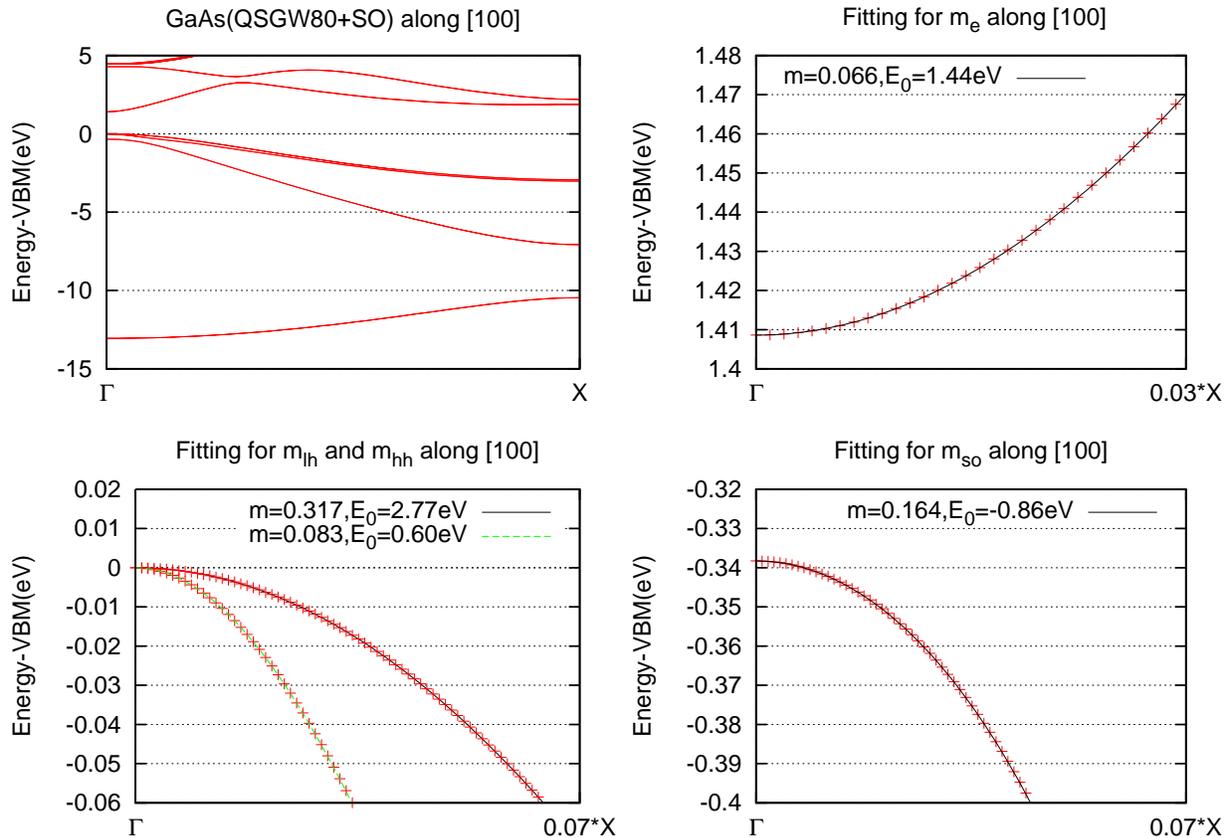}
\caption{\label{fig:massgaas}
Illustration about how to calculate effective masses for GaAs. 
The fitting window is between 0.01 and 0.05 eV for electrons 
(between -0.01 and -0.05 eV for holes) relative to the energy at the $\Gamma$ point. 
We use the fitting formula \cite{PhysRevB.35.7770}, 
$E(|\bfk|)(1+E(|\bfk|)/E_0)=\hbar^2|\bfk|^2/(2m)$, and 
determine the effective mass $m$ and correction $E_0$ to
the parabolic behavior. See text.}
\end{center}
\end{figure*}

\subsection{Effective mass of ZB-type semiconductors}
Effective mass plays a key role in the evaluation of transport properties and 
eigenvalues in quantum well structures and so on.
It is straightforward to calculate the effective mass in \QSGW\ in the ecalj package 
since we have interpolation procedure in the entire BZ; namely, without any extra
techniques such as Wannier interpolation, we can plot the band structure. 
We show the calculated effective mass along the [100] direction for ZB in
Table~\ref{tab:effmass} (the effective masses along other directions
are complicated \cite{chantis06a}).
As discussed in Ref. 40, 
the improvement of the prediction of the band-gap energy is essential 
for the correct prediction of the effective mass. 

In Fig. \ref{fig:massgaas}, we illustrate how to calculate the effective masses for GaAs.
To obtain effective masses along the [100] direction, we first calculate eigenvalues
(quasi-particle energies) along the [100] direction on a very dense ${\bf k}$ mesh. 
Then, for example, in the case of the electron mass $m_{\rm e}$,
we consider the electron branch at the bottom of the conduction band.
This branch of eigenvalues relative to the eigenvalue at the $\Gamma$ point 
can be represented by $E^{\rm e}(|{\bf k}|)$ as a function of the length of $|{\bf k}|$.
We now have $E^{\rm e}(|{\bf k}|)$ at dense $|{\bf k}|$ points along the [100] direction. 
Then, we perform a least square fitting for $E^{\rm e}(|{\bf k}|)$
in the energy window between 0.01 and 0.05 eV 
(or between -0.01 and -0.05 eV for holes).
This range corresponds approximately to room temperature. 
Here, we use the fitting formula 
$E^{\rm e}(|\bfk|)(1+E^{\rm e}(|\bfk|)/E^{\rm e}_0)=\hbar^2|\bfk|^2/(2m_e)$, 
where the two parameters $E^{\rm e}_0$ and $m_{\rm e}$ (effective mass) are
determined\cite{PhysRevB.35.7770}. $E^{\rm e}_0$ is a parameter that modifies a parabolic 
behavior slightly.
We apply the same procedure to other branches corresponding to
the heavy hole mass $m_{\rm hh}$, the light hole mass
$m_{\rm lh}$, and the mass for the split off band, $m_{\rm so}$. 
The calculated effective mass is summarized in Table~\ref{tab:effmass}.

As we see in Table~\ref{tab:effmass}, 
the agreement of theoretical effective masses 
with experiments ones is rather satisfactory,
especially for the electron mass. 
If we like to treat the subband structure of a superlattice such as CdS/CdSe
\cite{kocevski_first-principles_2015}, we have to use a method that 
can reproduce not only the band gap and band offset, but also the 
effective mass. This is never expected in methods such as the LDA and the
constant-shift procedure of the band gap (scissors operator procedure).
In this sense, 
careful treatment may be necessary to compare experimental 
data with the results in Ref. 39.

\section{Conclusions}
We have examined the ability of the \QSGW\ method and the hybrid method
\QSGW80 implemented in the ecalj package to predict band properties 
such as band gap energy, eigenvalues at special points, and effective mass. 
The ecalj package can be used easily 
since the parameters for calculation are automatically set. 
With the hybrid scheme \QSGW80, 
we can expect that the accuracy 
of band gaps can be $\sim$ 0.1 eV for usual semiconductors.
This level of accuracy is much higher than what we expect in the LDA.
Considering the fact that \QSGW\ can treat even metals accurately
\cite{jang_quasiparticle_2015,han_quasiparticle_2014,kotani_quasiparticle_2007}, 
we can expect that the \QSGW\ or hybrid \QSGW method can be applicable to 
complex systems such as metal/semiconductor interfaces.
However, it is necessary to know its limitations shown in this paper 
for such applications. 
In addition, note that \QSGW calculations are time-consuming, 
although we think there is so much room to improve the ecalj package for acceralating
the calculations. For typical cases, the values of computational time per QSGW iteration 
by using one node of Xeon (it contains 24 cores, Hitachi HA8000-tc/HT210 in Kyushu
university) are 2 min for C, 8 min for GaAs(ZB), 25 min for GaN(WZ), 36min for 
4H-SiC(8 atoms/unitcell), and 150 min for HgO(8 atoms/cell). 
Computational time depends not only
on the number of atoms per cell and the crystal symmetry but also on the atomic species. 
Usually, quasi-particle energies are converged within 0.01 eV after approximately 
five iterations.

\begin{acknowledgments}
This work was partly supported by the Advanced Low Carbon Technology
Research and Development Program (ALCA) of Japan Science and Technology
Agency (JST), and by Grants-in-Aid for Scientific Research 23104510 and 26286074. 
We also acknowledge the computing time provided by Computing System for
Research in Kyushu University. 
\end{acknowledgments}

\appendix
\section{How to reproduce the calculations in this paper by using the ecalj package}

The ecalj package is open for public use and all of the results presented 
in this paper can be reproduced by using the ecalj package by yourself. 
In this Appendix, the calculation procedure is briefly explained step by step. 
Owing to the limitation of space, we can show only the outline of the procedure. 
For detailed explanations and descriptions, visit 
https://github.com/tkotani/ecalj/\cite{ecalj}. 

\begin{enumerate}
\item {\bf Installation of ecalj:} The ecalj package can be downloaded and installed 
from Ref. 12.
The installer generates required binaries by using the Fortran compiler 
and performs minimum test calculations successively.

\item {\bf Preparation of {\tt ctrls} file:} To start a calculation of target material, 
we need to prepare a {\tt ctrls.*} file, which contains information on the crystal structure.
For the extension {\tt *}, we usually use the name of the material, such as {\tt gaas, zno}, 
and so on. The {\tt ctrls} file is explained in detail at the link\cite{ecalj}. 
Instead of obtaining {\tt ctrls.*} manually, you can use a converter, for example, 
{\tt vasp2ctrl} is prepared for extracting information on the crystal structure from 
the POSCAR file for VASP (concerning VASP, see {\tt https://www.vasp.at/}). 

\item{\bf Generating {\tt ctrl} file:} From the {\tt ctrls.*}, next we generate a {\tt ctrl.*} file by using a 
python script, {\tt ctrlgenM1.py}, included in the ecalj package. 
In addition to the information on the crystal structure, 
{\tt ctrl.*} contains all of the parameters that control the calculations, 
such as information on MTOs, xc potential, number of k-points, relativistic 
treatment, and so on. 
In the generated {\tt ctrl.*}, default settings are set to guarantee a reasonable 
calculation, but we need to edit them depending on the necessity. 
(With {\tt ctrl.*}, we can perform DFT calculations.)

\item{\bf Generating {\tt GWinput}:} In addition to the {\tt ctrl} file, we need 
one more input file, {\tt GWinput}, which 
can be generated from {\tt ctrls.*} by using a shell script 
({\tt mkGWIN\_lmf2} also included in the {\tt ecalj} package). 
{\tt GWinput} contains settings for GW calculation.

The input files used in this paper are basically produced 
from {\tt ctrls.*} by following the above processes (2)-(4). 
To facilitate the reproduction of the present calculations by the users, 
we packed {\tt ctrls.*}, {\tt ctrl.*}, and {\tt GWinput} for each material, 
which are available as supplementary data of this article at the web page\cite{supplement}. 

\item{\bf Performing {\rm QSGW}:} With the two input files {\tt ctrl.*} and {\tt GWinput}, 
we can perform \QSGW\ calculations by using a script {\tt gwsc}. 
We set the number of iterations when we submit a job of {\tt gwsc}. 
The output files of \QSGW\ calculations are stored in {\tt rst.*} and {\tt sigm.*}. 
The outputs for the materials treated in this paper can be found also at the web 
page\cite{supplement}. 
After we examine the convergence of eigenvalues, we proceed with post processing 
such as plotting the energy bands. 

\item{\bf Postprocessing:} 
Once the self-consistency is obtained, 
we can extract any material properties from the calculated electronic structure.
How to do it is not trivial and an additional calculation or postprocessing is needed. 
As concerned with this paper, it is necessary to plot and fit energy bands for estimating 
the effective mass. 
As an example, the energy band of CdO is shown in Fig.~\ref{fig:app}
The energy-band plots for the other materials and fitting results can be found 
at the web page\cite{supplement}. 
The files {\tt syml.*}, which describes the k-point path for the band structure plot, 
are also packed. 
More details concerning the band plot 
can be found in the README given in the page\cite{supplement}. 

\begin{figure}[h]
\begin{center}
\includegraphics[width=8cm]{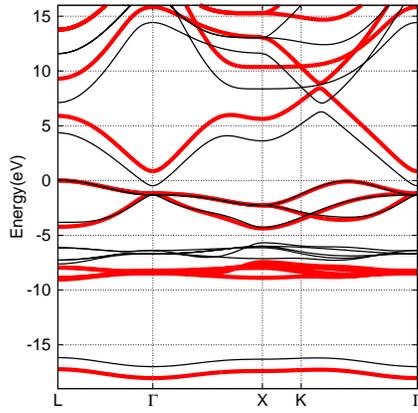}
\caption{\label{fig:app}
Calculated band structure of CdO 
by using the LDA (thin black lines) and \QSGW80 (thick red lines). 
The energy is relative to the valence band maximum.}
\end{center}
\end{figure}

\end{enumerate}

\bibliographystyle{jjap}

\end{document}